\documentclass[a4paper]{jpconf}
\usepackage{graphicx}
\usepackage{amssymb}
\def\be{\begin{equation}}
\def\ee{\end{equation}}
\begin{document}
\title{Neutrino masses and mixing}

\author{Alexei Yu. Smirnov\footnote{Invited talk given at TAUP2005, 
September 10 - 14, 2005, Zaragoza, Spain.}}

\address{International Centre for Theoretical Physics 
Strada Costiera 11, 34014 Trieste, Italy\\
Institite for Nuclear Research, RAS, Moscow, Russia}

\ead{smirnov@ictp.it}

\begin{abstract}
Status of determination of the neutrino masses and mixing is formulated 
and possible uncertainties, especially due to presence of the sterile neutrinos,  
are discussed. The data hint an  existence of special 
``neutrino'' symmetries. If not accidental these symmetries 
have profound implications and can substantially change the unification program. 
The key issue on the way to underlying physics is relations between 
quarks and leptons.  The approximate quark-lepton symmetry or universality can be 
reconciled with strongly  different patterns of masses and mixings due to nearly 
singular character of the mass matrices or screening of the Dirac structures 
in the double see-saw mechanism.   
\end{abstract}

\section{Introduction}

In the first approximations the pattern of lepton mixing 
has been  established.   
The 2-3 mixing is consistent with maximal,  1-2 mixing is large but not maximal  and 
1-3 mixing is small and consistent with zero. 
The next step is  
determination of  detailed structure 
of the mass and mixing, in particular,  measurements of the {\it deviations} of 
2-3 mixing from maximal and 1-3 mixing -  from zero.  

There are two key issues on the way to the  underlying physics: 
 
\begin{itemize}

\item 
possible existence of  new ``neutrino'' symmetries 
behind the pattern of neutrino mass and mixing;

\item
relation between quarks and leptons -   
their possible symmetries and unification. 

\end{itemize}

The two questions are related: 
establishing specific symmetry in the neutrino sector 
may substantially change the unification program.

\section{Results and  uncertainties}

\subsection{Summarizing results} 

Masses: The solar and the atmospheric mass differences squared 
give the lower bound on ratio of the second and third masses: 
\be
\frac{m_2}{m_3} \geq \sqrt{\frac{\Delta m_{21}^2}{\Delta m_{31}^2}} = 0.15 - 0.20.
\label{ratiom}
\ee
Both cosmology and the double beta decay probe the 
sub-eV region which corresponds to the quasi-degenerate mass spectrum 
giving the upper bound $m < 0.2 - 0.4$ eV \cite{cos}.  
If the Heidelberg-Moscow result~\cite{hm} is confirmed 
and if it is due to exchange of the light Majorana neutrinos,  the 
neutrino mass spectrum should be strongly degenerate.

\begin{figure}[t]
    \begin{tabular}{cc}
      \resizebox{0.50\textwidth}{!}{\includegraphics{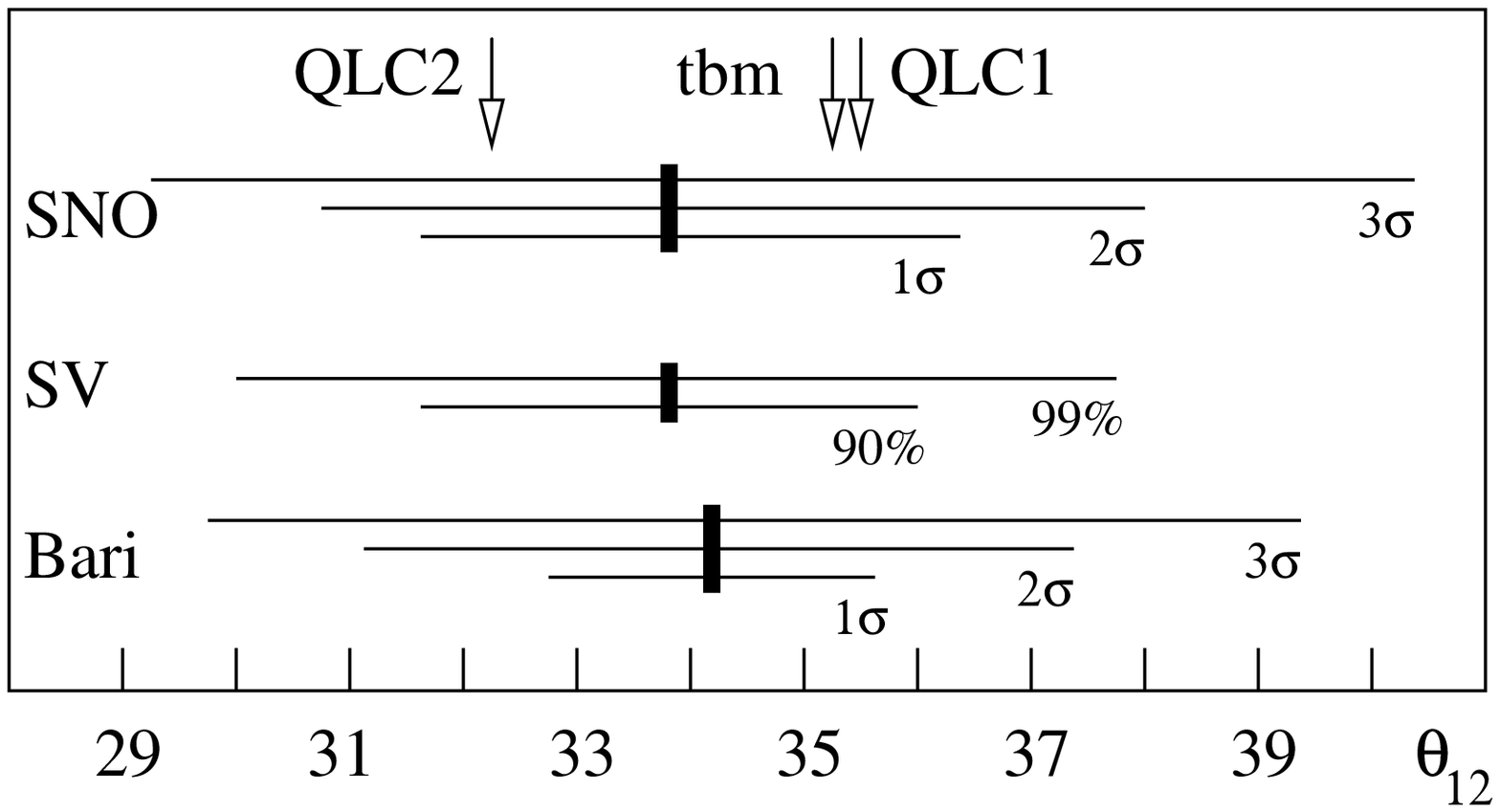}} &
      \hspace{0.3cm}
      \resizebox{0.42\textwidth}{!}{\includegraphics{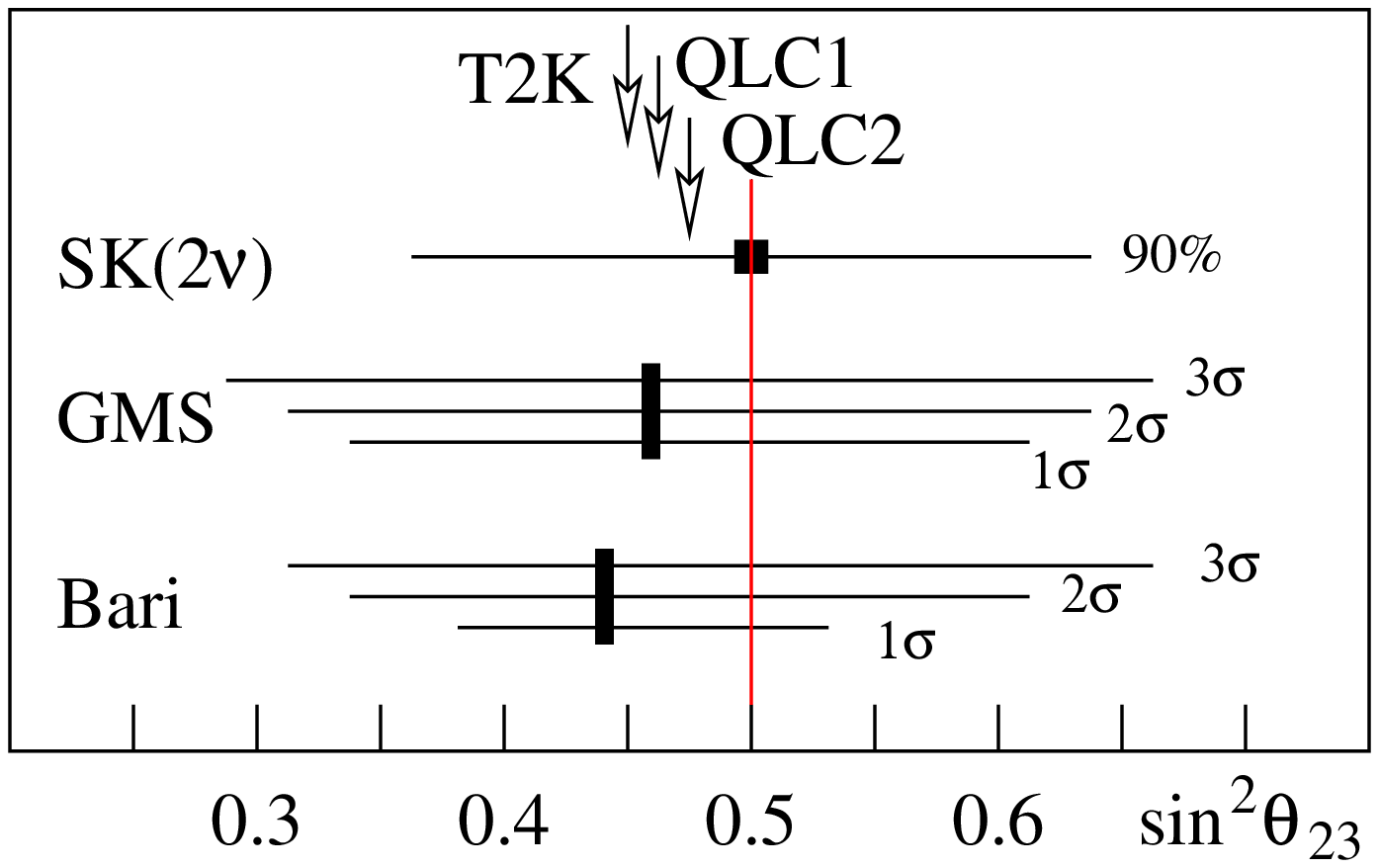}} \\
      (a) & (b)
    \end{tabular}
    \caption{The best fit values and 
the allowed regions of lepton mixing angles 
at different confidence levels determined by different groups.  
\textbf{(a)} $\theta_{12}$ from SNO~\cite{sno}, SV~\cite{sv} and 
Bari~\cite{bari}. Shown are predictions from QLC and tri-bimaximal mixing.  
\textbf{(b)} $\sin^2 \theta_{23}$ from  SK~\cite{atm}, GMS~\cite{concha}, 
Bari~\cite{bari}. Shown are expectations from QLC and sensitivity limit of T2K 
experiment~\cite{T2K}.}
  \label{12mix}
\end{figure}

\begin{figure}[t]
\begin{center}
\resizebox{0.54\textwidth}{!}{\includegraphics{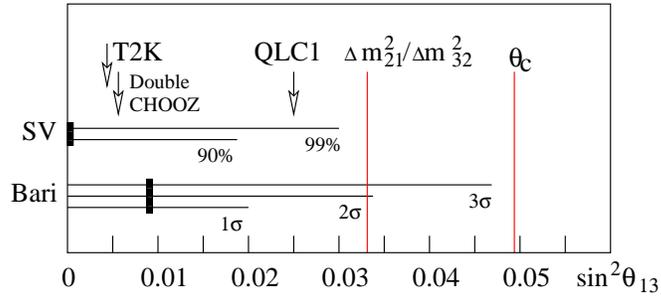}} 
\end{center}
\caption{The best fit values and
the allowed regions of lepton mixing angle $\theta_{13}$
at different confidence levels determined by different groups: 
SV~\cite{sv} and Bari~\cite{bari}. Shown are also some predictions and 
the sensitivity limits 
of  Double CHOOZ~\cite{DC} and T2K~\cite{T2K}.}
\label{13mix}
\end{figure}

Results  on determination of 
the lepton mixings are summarized in fig. \ref{12mix}, \ref{13mix}.

1).  1-2 mixing: 
there is a very good agreement of central values  and
reasonable agreement of the allowed ranges at different confidence
levels obtained by three different groups \cite{sno,sv,bari}.
The $3\nu$ analysis does not change the best fit value of mixing 
in comparison with 2 neutrino analysis but the error 
bars become smaller.

After the recent  SNO publication~\cite{sno}, the 1-2 mixing has  
shifted to larger values by $\Delta \theta_{12} \sim 1.6^{\circ}$ 
due to increase of the CC/NC ratio  
and now the b.f. value equals
$
\theta_{12} = 33.9^{\circ}. 
$
In fig. \ref{12mix}a  we show also several theoretical benchmarks:
predictions from 
(i) from QLC1 scenario (bi-maximal mixing from the
neutrino sector, see sec. 4.3): 
$\theta_{12}  = 35.4^{\circ}$, (ii) the QLC2 scenario:
$\theta_{12} =  
45^{\circ} - \theta_C \approx  32.2^{\circ}$
(the bi-maximal mixing from the charge lepton sector),
and (iii) tri-bimaximal mixing $\sin \theta_{12} = 1/\sqrt{3}$,
or $\theta_{12}  = 35.2^{\circ}$. 
All three predictions are within $1\sigma$.
Predictions from the tri-bimaximal mixing and  QLC1
almost coincide, the b.f. value is in between
the QLC2 and two other predictions.
To disentangle these two possibilities
one needs to measure the 1-2 mixing with accuracy
$\Delta \theta_{12} \sim  1^{\circ}$ or
$\Delta \sin^2 \theta_{12} \sim  0.015$ ($5\%$).\\

2). The 2-3 mixing is in agreement with  maximal one (fig.~\ref{12mix}b).
A shift from maximal mixing has been found when
effects of 1-2 sector have been included 
in the analysis \cite{orl}. According to \cite{concha}
$\sin^2\theta_{23} = 0.47$ and slightly larger
shift, $\sin^2\theta_{23} = 0.44$,  follows from the analysis \cite{bari}. 
So, the deviation from maximal mixing can be quantified as   
\be
D_{23} \equiv 0.5 - \sin^2\theta_{23} \sim 0.03 - 0.06. 
\ee
The shift is related to the  excess of 
e-like atmospheric neutrino events in the sub-GeV range detected by 
SuperKamiokande (SK).  
The excess is proportional to the deviation $\Delta N_e/N_e \propto D_{23}$ 
\cite{orl}. 
No shift from maximal mixing has been found in the
recent SK $3\nu$-analysis even after inclusion 
of the 1-2 sector (LMA oscillations) \cite{suzuki}. The difference of results may be 
related to  treatment of uncertainties in the atmospheric neutrino flux
normalization. The change of normalization competes with
the effect of LMA oscillations. In the SK analysis
the excess of e-like events is explained completely 
by the normalization. In the analyses
\cite{concha,bari} the excess is explained by the normalization
partially, since certain distribution of  the normalization factors is assumed.

Still large deviation from maximal mixing is allowed:
\be
D_{23}/\sin^2\theta_{23} \sim 0.4 ~~~(2\sigma). 
\ee

3). The 1-3 mixing is consistent with zero 
(fig. \ref{13mix}). Small non-zero best fit value from the analysis
\cite{bari} is related to the  angular dependence 
of the multi-GeV e-like events measured by SuperKamiokande.
The most conservative $3\sigma$ bound is $\sin^2 \theta_{13} < 0.048$
\cite{bari}. There are several benchmarks here:
A very appealing possibility, $\theta_{13} = \theta_{C}$,  
seems to be excluded at more than  $3\sigma$ level. The ratio of the
solar and atmospheric neutrino mass scales, 
\be
\sin^2 \theta_{13} = r = \frac{\Delta m_{21}^2}{\Delta m_{31}^2} = 0.033, 
\ee
is allowed at about $2\sigma$ level. An 
additional (model dependent) factor of the order 0.3 - 2   
may appear in this relation. 
Much smaller values of $\sin^2 \theta_{13}$ would
imply most probably certain symmetry of the mass matrix.

There are several lower bounds on the 1-3 mixing:
(i) even if equality  $\sin^2 \theta_{13} = 0$ holds  
at some high energy scale, $\Lambda$ (presumably GUT or scale of flavor 
physics), a nonzero value of the order  $\sin^2 \theta_{13} = 0.003$
is generated due to the renormalization group effect (unless
some accidental cancellation occurs) \cite{manf}.
(ii) Smaller values, $\sin^2 \theta_{13} = 10^{-4}$,  
are expected due to possible contributions to the mass matrix from the Planck scale 
interactions  $\sim v_{EW}^2/ M_{PL}$ \cite{vv}.

\subsection{Uncertainties}

Three types of possible effects can influence interpretation of
the neutrino results.  

1). Existence of new neutrino states -
sterile neutrinos which  are of great interest not only in 
connection to the LSND result.  
If these states are light they can directly (dynamically)
influence observations. 
If sterile neutrinos  are heavy and decouple from the low 
energy physics, they may substantially change implications of
the results for the fundamental theory.

2). Presence of the non-standard (short range) neutrino interactions
can change values of the extracted neutrino parameters.

3). Interactions with hypothetical light scalar fields  produce 
''soft'' neutrino masses  
which depend on properties of medium. These masses may change with 
time and be related to the dark energy in the universe \cite{mavan}.

At present, however there is no well established results which
could testify for deviations from the ``standard'' $3\nu$ mixing scheme  and
the standard matter interactions.\\

Let us consider one  
aspect of possible presence  of sterile neutrinos - ambiguity 
in interpretation of the neutrino 
results. Even small mixing of active neutrinos with sterile ones 
can substantially change the structure of active neutrino mass matrix, 
in particular, inducing large mixing \cite{abdel}.
Suppose the active neutrinos acquire (e.g., via seesaw)
the Majorana mass matrix $m_a$. Consider one sterile neutrino, 
$S$,   with Majorana mass $M$
and mixing masses with active neutrinos
$m_{iS}$ ($i = e, \mu, \tau$). If $M \gg m_{iS}$,  
then after decoupling of $S$ the mass matrix of active neutrinos becomes 
\be
(m_{\nu})_{ij} = (m_a)_{ij} - m_{iS}m_{jS}/M, 
\ee
where the last term is the matrix induced by $S$. 
New neutrino states are irrelevant if
\be
m_{iS}m_{jS}/M \ll (m_a)_{ij}.  
\label{cond}
\ee

The smallest matrix elements are in the case 
of normal mass hierarchy ($m_1 \approx 0$).
The data can be well described by  
\begin{equation}
m_{\nu} =
\frac{m_3}{2} 
\left(\begin{array}{ccc}
0 & 0 & 0\\
0 & 1 & - 1\\
0 & - 1 & 1
\end{array}
\right) +
\frac{m_2}{3}
\left(\begin{array}{ccc}
1 & 1 & 1\\
1 & 1 & 1\\
1 & 1 & 1
\end{array}
\right) 
\label{tribi}
\end{equation}
(which by the way, corresponds to the tri-bimaximal mixing), 
where $m_2 = \sqrt{\Delta m_{21}^2}$ and $m_3 = \sqrt{\Delta 
m_{31}^2}$.
Assuming  flavor ``blindness'':  $m_{iS} = m_S$,  we can rewrite 
the condition (\ref{cond}) using  the smallest elements in (\ref{tribi}) as
$
m_S^2/M \ll m_2/3, 
$
or
\be
\sin\theta_S^2~ M \ll m_2/3 \sim 3\cdot 10^{-3} ~{\rm eV}. 
\label{smixing}
\ee
Here $\theta_S$ is the active-sterile mixing angle:
$\sin\theta_S = m_S/M$.

For $M \sim 1$ eV we obtain from (\ref{smixing})
$\sin^2\theta_S  \ll 3 \cdot 10^{-3}$. This means that if the LSND 
interpretation as an effect of additional neutrino states is confirmed,
its impact on the neutrino mass matrix is strong and can not
be considered as perturbation.
If the effect is not confirmed, the MiniBOONE sensitivity
is not enough to exclude strong effect of new states.
For $M \sim 1$ MeV we get $\sin\theta_S^2  < 10^{-9}$. 

\section{Neutrino symmetry}

Several observations may testify for special symmetry(ies)
associated to neutrinos:  
(i) maximal (close to maximal) 2-3 mixing; 
(ii) zero (very small) 1-3 mixing; 
(iii) special values of 1-2 mixing; 
(iv) degenerate mass spectrum; 
(v) hierarchy of mass squared differences. 
%
Some of these features can originate from the same underlying symmetry.\\


\subsection{Schemes of mixing}
In connection to the above observations the following schemes of mixing
can be of relevance. 

1). The bi-maximal mixing \cite{bim}:
\be
U_{bm} = U_{23}^m U_{12}^m =  
\frac{1}{2}
\left(\begin{array}{ccc}
\sqrt{2} & \sqrt{2} & 0\\
-1 & 1 & \sqrt{2}\\
1 & - 1 & \sqrt{2}
\end{array}
\right). 
\label{bimax}
\ee
Identification $U_{PMNS} = U_{bm}$ is not possible due to
strong  (5 - 6) $\sigma$ deviation of the 1-2  mixing from
maximal. However,  $U_{bm}$ can play a role of dominant structure
or matrix in the lowest order. 
Correction can originate from the charged lepton
sector (mass matrix), so that
$U_{PMNS} = U'U_{bm}$ and   in analogy with quark mixing 
$U' \approx U_{12}(\theta_C)$. 
It generates simultaneously 
deviation of the 1-2 mixing from maximal and non-zero
1-3 mixing, which are related.  

2). Tri-bimaximal mixing \cite{tbm} 
\be
U_{tbm} = U_{23}^m U_{12}(\theta_{12}) =  
\frac{1}{\sqrt{6}}
\left(\begin{array}{ccc}
2 & \sqrt{2} & 0\\
-1 & \sqrt{2} & \sqrt{3}\\
 1 & - \sqrt{2} & \sqrt{3}
\end{array}
\right), 
\label{bimax}
\ee
where $\sin^2 \theta_{12} = 1/3$.  
Here $\nu_2$ is tri-maximally mixed: 
in the middle column three flavors mix maximally,
whereas  $\nu_3$ (third column) is bi-maximally mixed.
This matrix is in a  good agreement with data, 
in particular, $\sin^2\theta_{12}$ is close to the
present best fit value 0.31.


\subsection{$\nu_\mu - \nu_\tau$ symmetry}
Maximal 2-3 mixing and zero 1-3 mixing
can be consequences of the
$\nu_{\mu} - \nu_{\tau}$ permutation symmetry of the neutrino mass matrix \cite{mutau}.
General form of such a matrix in the flavor basis is
\be
M = 
\left(\begin{array}{ccc}
A & B & B\\
B & C &  D\\
B & D & C
\end{array}
\right). 
\label{23mat}
\ee
The permutation symmetry can be  a part of {\it e.g.} discrete $S_3$ or $D_4$  
groups.

The problem is that the symmetry is broken for charged
leptons since $m_{\mu} \ll m_{\tau}$.  
So, it can not be the  symmetry of complete theory. 

In principle the symmetry violation can be  weak -   
characterized  by small parameter $m_{\mu}/m_{\tau}$.  
In this connection one can consider an
example, when the charged lepton mass matrix is
``democratic'' (with all elements to be equal each other) and the
neutrino mass matrix is diagonal \cite{zzz}. These  matrices
(particular cases of (\ref{23mat})) lead to hierarchical
charge lepton spectrum  (in fact with only one non-zero eigenvalue)
and degenerate spectrum of neutrinos. 
Diagonalization of the charged lepton mass matrix leads
to maximal 1-2 mixing and non-maximal 2-3 mixing $\sin^2 2 \theta_{23}
= 8/9$. So, small corrections
should be introduced to generate masses of muon and electron,
as well as to produce mass split of neutrinos and to correct mixings.

Let us discuss possible solutions of the problem with 
$\mu - \tau$ symmetry. 
The symmetry can be broken spontaneously,  and for this the
extended Higg's sector is required. Several possibilities exist 
to explain why it shows up in the neutrino sector only. 
Apparently the difference of neutrinos and charged leptons should be related to their 
RH components.   

1).  Auxiliary symmetry, e.g. $Z_2$,  can be
introduced  to protect neutrino sector 
from $\mu - \tau$ breaking. 
$l_R$ and $\nu_R$ should have
different properties with respect to this auxiliary symmetry.

2). The symmetry basis can differ from the flavor basis.
So, in the symmetry basis 
(one should speak about 2 - 3 permutations)
the mass matrix of charged leptons is off-diagonal. 

3). Among other flavor symmetries  $A_4$  looks very appealing 
\cite{a4}. It has one triplet representation and three different singlet 
representations, ${\bf 1},~ 
{\bf 1'},~ {\bf 1''}$,  which provides with enough freedom to explain data.  
Three leptonic doublets
form the triplet of $A_4$: $L_i = (\nu_i, l_i)  \sim {\bf  3}$,
$i = 1, 2, 3$. 
Required lepton mixing is generated due to different 
$A_4$ transformation properties 
of the right handed components of charged leptons and neutrinos.  
In some models: $l^c_i \sim {\bf 1},~ {\bf 1'},~ {\bf 1''}$, whereas 
$N^c_i \sim {\bf 3}$.  In  other models {\it vice versa}: $l^c_i \sim {\bf 3}$, 
$N^c_i \sim {\bf 1},~ {\bf 1'},~ {\bf 1''}$. 

4). See-saw induced symmetries. 
In this case neither Dirac mass matrix  nor Majorana mass matrix
of the RH neutrinos have  the  required symmetry  but they have certain
structures (hierarchies of matrix elements).
The symmetry appears as a result of the see-saw mechanism. 

In specific models  some combinations 
of  these mechanisms are realized \cite{model}.

\subsection{Real or accidental}
The main question here is   
whether the ``neutrino'' symmetries are accidental or real, 
that is,  have some  physics behind. 
Models proposed so far are rather complicated with a
number of  {\it ad hoc} assumptions.
It is difficult to include quarks in these models.
Further unification looks rather problematic.
Asymmetries between  neutrinos and leptons 
are embedded into theory from the beginning.
This shows the price one should to pay for realization 
of the symmetries.


Furthermore,  the facts behind the symmetries -
maximal 2-3 mixing and relatively small 1-3 mixing are not yet
well established. Still significant deviation
of 2-3 mixing is possible and 1-3 mixing can be not so small.
Structure of the neutrino mass matrix depends substantially
on these deviations.  
So, it may happen that symmetry constructions are
simply misleading.

On the other hand if symmetries are not accidental,
they have consequences of the fundamental
importance as the models constructed show.
New structures and particles are predicted,  unification
path may differ substantially from what we
are considering now,  etc.. The symmetries may give some clue
for understanding fermion masses in general.

The key question is how to test this?
Obviously, we need to search for and  
measure deviations: of 2-3 mixing from maximal, $D_{23}$, and 1-3 mixing,   
$\sin \theta_{13}$, from zero.  
In the context of specific models
the deviations (though small) are expected anyway.  
The facts we are discussing
can originate from the same symmetry and violation of this symmetry
will lead then to relations between $D_{23}$ and $\sin \theta_{13}$. \\

It may happen that the symmetries are not accidental  
but the underlying theory has not been found yet. 
In this connection let us come back to the issue of active-sterile mixing.
Let us assume that the couplings of $S$ with active neutrinos are universal:
\be
m_{iS} = m_S (1, 1, 1) = m_2/\sqrt{3}. 
\ee
Then the induced matrix has form:
$
m_{ind} = m_2 D/3, 
$
where $D$ is the democratic matrix (the second 
matrix in (\ref{tribi}). 
Suppose that the original active neutrino mass matrix has structure 
of the first matrix in (\ref{tribi}). 
Then the sum,  $m_{\nu} = m_a + m_{ind}$,  reproduces the mass
matrix for the tri-bimaximal mixing (\ref{tribi}).
With two sterile neutrinos whole structure (\ref{tribi})
can be obtained.  

Clearly this possibility changes implications of the neutrino results.  
Since $S$ is beyond the SM structure 
extended with RH neutrinos,  it may be easier to realize ``neutrino'' symmetries as 
a consequence of certain  
symmetry of its couplings with active neutrinos.

\section{Leptons and Quarks}


\subsection{Comparing leptons and quarks}

There is an apparent correspondence between quarks and leptons. 
Each quark has its own counterpartner in the leptonic sector.  
Leptons can be treated as the 4th color \cite{pati}
following the Pati-Salam $SU(4)$ unification symmetry.
Unification  is possible, 
so that quarks and leptons form multiplets of the extended gauge group.
The most appealing one is SO(10) \cite{so10}, where all known components of quarks and 
leptons
(including the RH neutrinos) form unique 16-plet.
It is difficult to believe that all these features are accidental.
Though it is not excluded that the
quark-lepton connection has some more complicated form, 
e.g., of the quark - lepton complementarity \cite{qlc,qlc1}. 
\begin{figure}[t]
\begin{center}
\resizebox{0.47\textwidth}{!}{\includegraphics{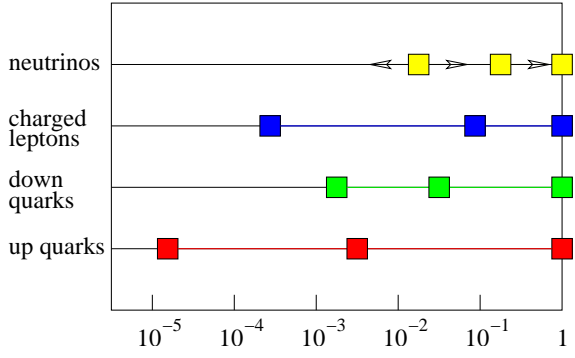}} 
\end{center}
\caption{Mass hierarchies of quarks and leptons.  
The mass of heaviest fermion of a given type is taken to be 1.}
  \label{ratios}
\end{figure}

In the quark sector we have rather complete information about
masses and mixings and still no explanation has been found. 
It seems,  neutrinos have not helped yet, and on the contrary, 
made a situation even more complicated. 
Comparison of masses and mixing in the quark and lepton
sectors and establishing certain relations between them 
may give some insight.

Apparently the mixing patterns of leptons and quarks is strongly different:
The only common  feature is that the 1-3 mixing
(between the ``remote'' generations) is small in both
cases. Two other angles are not equal but complementary in a sense
that they sum up to maximal mixing:
\be
\theta_{12} + \theta_C  = \frac{\pi}{4},  
\label{qlcrel}
\ee
and similar approximate relation can  be written for the 2-3 mixings.
For various reasons it is difficult to  expect precise relation but
qualitatively one can say that,  the 2-3 mixing in the lepton sector
is close to maximal  because the corresponding quark mixing is
small, the 1-2 mixing deviates from maximal substantially because
the 1-2 (Cabibbo) quark mixing is relatively large.
It seems that for the third angle we do not expect simple relation
and apparently the quark feature 
$\theta_{13} \sim \theta_{12} \times \theta_{23}$
does not work in the lepton sector.

The ratio of neutrino masses (\ref{ratiom})
can be compared with ratios for charged leptons and quarks (at $m_Z$
scale):
$m_\mu/m_\tau =  0.06$, $m_s/m_b = 0.02 - 0.03$,  $m_c/m_t = 0.005$.
The neutrino hierarchy - see eq. (\ref{ratiom})  (if exists at all -  still the 
degenerate spectrum is not excluded) is the weakest one.
This is consistent with possible mass-mixing relation: 
large mixings are associated to weak mass  hierarchy.

In fig.~\ref{ratios} we show the mass ratios for three generations.
The strongest hierarchy and geometric relation $m_u \times m_t \sim m_c^2$
exist for the upper quarks. Apart from that  no simple relations show up.
What is behind this picture? Symmetry, regularities, relation?
In the quark sector we can speak about fermion families with
weak interfamily connection (mixing)  which means strong flavor 
alignment. In the lepton sector the alignment is weaker.  
Furthermore, peculiar situation with fermion masses 
is that spectra have small number 
of states (levels) - 3,  and on the other hand
there is no simple relations between parameters of spectra.
It looks like the observed pattern  is an interplay of some
regularities and randomness (``anarchy'').

\subsection{Quark-lepton symmetry and Quark-lepton universality}

The picture described in the previous section is still  
consistent  with the approximate quark-lepton symmetry or universality. 
However, the symmetry is realized in terms of mass matrices (matrices of the
Yukawa couplings) and not in terms of observables - mass ratios and
mixing angles. 

The key point is that similar mass matrices can lead to
substantially different mixing angles and masses (eigenvalues)
if the matrices are nearly singular (rank-1) \cite{sing,dors}. 
The singular matrices are ``unstable''
in a sense that small perturbations can lead to strong variations of
mass ratios and mixing angles (the latter -  in the context of seesaw.

Let us consider the universal structure for the mass matrices
of all quarks and leptons \cite{dors}:
\be
Y_u \sim Y_d \sim Y_D \sim Y_M \sim Y_L \sim Y_0, 
\ee
where  $Y_D$ is the Dirac type neutrino Yukawa matrix,
$Y_M$ is the Majorana type matrix for the RH neutrinos
and  $Y_0$ is the singular matrix. As an important example we can take
\be
Y_0 =  
\left(\begin{array}{ccc}
\lambda^4 & \lambda^3 & \lambda^2\\
\lambda^3 & \lambda^2 & \lambda\\
\lambda^2 & \lambda & 1
\end{array}
\right), ~~~~ \lambda \sim 0.2 - 0.3.
\label{anz}
\ee
This matrix has only one non-zero eigenvalue and determines 
mixing angles for the third generation.

Let us introduce perturbations  $\epsilon$ in the following form
\be
Y^f_{ij} = Y^0_{ij} (1 + \epsilon_{ij}^f), ~~~ f = u, d, e, \nu, N ,
\label{pert}
\ee
where $Y^0_{ij}$ is the element of the original singular matrix.
This form can be justified, {\it  e.g.} in the context of the Froggatt-Nielsen
mechanism~\cite{fn}. (The key element is the form of perturbations (\ref{pert})
which distinguishes the ansatz (\ref{anz}) from other possible schemes with 
singular matrices.) 
It has been shown that small perturbations
$\epsilon \leq 0.25$ are enough  to explain large difference in mass hierarchies
and mixings of quarks and leptons \cite{dors}.

Smallness of neutrino mass is explained by the seesaw mechanism.
Furthermore, nearly singular matrix of the RH neutrinos leads
to an enhancement of the lepton mixing~\cite{ssenh} and to flip of sign of mixing
angle which comes from diagonalization of the neutrino mass matrix.
So, the angles from the charged leptons and neutrinos sum up, 
whereas in quark sector mixing angles from up and down quark mass matrices
subtract.\\

Keeping this in mind one can consider the following
``working'' hypothesis:

\noindent
1). No particular ``neutrino'' symmetry exists,
and in general  one expects some
deviation of the 2-3 mixing from maximal
as well as non-zero 1-3 mixing.
Nearly maximal 2 -3 mixing would be accidental in this case.

\noindent
2). Seesaw mechanism with the scale of RH neutrino masses
$M \sim 10^7 - 10^{15}$ GeV explains smallness of neutrino mass.
The upper part of this range is close to the GU scale and can be
considered as
indication of the Grand Unification.

\noindent
3). The quark-lepton unification or Grand Unification are realized in some 
form,  e.g. $SO(10)$.

\noindent
4). The quark-lepton symmetry is (weakly) broken and 
there are some observable consequences like $m_b = m_\tau$.

\noindent
5). Large lepton mixing is a consequence of the seesaw type-I mechanism
 - the seesaw enhancement of lepton mixing
due to special structure of the RH neutrino mass matrix,
(or/and of the contribution from the type II seesaw).

\noindent
6). Flavor symmetry or/and physics of extra dimensions determine
this special structure.

\subsection{Quark-lepton complementarity (QLC)}

Being confirmed the complementarity (\ref{qlcrel}) would require
certain modification of the  picture described above \cite{qlc}. 
The latest determination of the solar mixing angle gives
$
\theta_{12} + \theta_C = 46.7^{\circ} \pm 2.4^{\circ}  ~~~(1\sigma)
$
which is consistent with maximal mixing angle within $1\sigma$.
Is the QLC-relation accidental or there is some physics behind, 
that should include non-trivial quark-lepton connection?
The fact that for the 2-3 mixings the approximate complementarity is also 
fulfilled hints some more serious  reasons than just numerical 
coincidence.

A general scheme is that
\be
``{\rm lepton~ mixing} =  {\rm bi-maximal~mixing} - {\rm CKM}''. 
\ee
There is a number of non-trivial conditions for the exact QLC relation
to be realized.

(i) Order of rotations: apparently $U_{12}^m$ and $U_{12}^{CKM \dagger}$
should be attached 
\be
U_{PMNS} \equiv U_L^{\dagger}U_{\nu} = ...U_{23}^m ... U_{12}^m
U_{12}^{CKM \dagger}
\label{order}
\ee
(two last rotations can be permuted). Different order leads to corrections
to the exact QLC relation;
(ii) Matrix with CP violating phases should not appear between
$U_{12}^{CKM \dagger}$ and $U_{12}^m$; 
(iii) Presumably the quark-lepton symmetry 
which leads to the QLC relation is realized at high mass scales.
Therefore the renormalization group effects should be small enough,
etc..

Let us describe two possible scenarios which differ by  origin
of the bi-maximal mixing and
lead to different predictions.

1). QLC1: The bi-maximal mixing is generated by the neutrino
mass matrix, presumably due to  seesaw. The charged lepton mass matrix
produces  the CKM mixing as a consequence of the q-l symmetry:
$m_l = m_d$. In this case the order of matrices (\ref{order})
is not realized ($U_{12}^{CKM}$ should be permuted with $U_{23}^m$), 
and consequently the QLC relation is modified:
\be
\sin \theta_{12} = \sin (\pi/4 -\theta_C) + 0.5\sin \theta_C (\sqrt{2} -1). 
\label{qlc1}
\ee
Numerically we find $\tan^2\theta_{12} = 0.495$ which
is practically indistinguishable from the tri-bimaximal mixing (fig.~\ref{12mix}a).

2). QLC2: Maximal mixing comes from the charged lepton mass matrix
and the CKM mixing originates from the neutrino mass matrix due to
the q-l symmetry: $m_D \sim m_u$ (assuming also that in the context of seesaw
the RH neutrino mass matrix does not influence 
mixing). In this case  the QLC relation is satisfied precisely:
$\sin \theta_{12} = \sin (\pi/4 -\theta_C)$. 

There are two main issues related to the QLC relation:

(1) origin of the bi-maximal mixing; 

(2) mechanism of propagation  of the CKM mixing 
from the quark to the lepton sector.
The problem here is large difference of mass ratios
in the quark and lepton sectors:  $m_e/m_\mu = 0.0047$, 
$m_d/m_s = 0.04 - 0.06$,  as well as difference of masses of muon and
s-quark at the GU scale. 
This means that mixing should weakly depend or be independent on masses. 

Mass matrices are different for quarks and leptons
and ``propagation'' of the CKM mixing leads to corrections
to the QLC relation of the order~\cite{qlc1}
\be
\Delta \theta_{12} \sim \theta_C m_d/m_s \sim 0.5 - 1.0^{\circ}. 
\ee

The Cabibbo mixing can be transmitted to the lepton sector in
more complicated way (than via the q-l symmetry).
In fact, $\sin \theta_C$ may turn out to be the generic parameter
of theory of  fermion masses and therefore to appear
in various places: mass ratios, mixing angles.
The relation:
$\sin \theta_C \approx \sqrt{{m_{\mu}}/{m_{\tau}}}$ 
is in favor of this possibility.

So, if not accidental the QLC relation may have two different
implications:
One includes the  quark-lepton symmetry, existence of some additional
structure which produces the bi-maximal mixing, weak dependence of the
mixings on mass eigenvalues. 
Alternatively, it may imply certain flavor physics with
$\sin \theta_C$ being  the ``quantum'' of this physics.

\subsection{Screening of Dirac structure}

The quark -lepton symmetry manifests  as certain relation
(similarity) between the Dirac mass matrices of quarks and leptons,
and it is this  feature which creates problem for explanation of
strongly different mixings and possible existence of the ``neutrino'' symmetries.
Let us  consider an extreme case when in spite of the q-l unification, 
the Dirac structure in the lepton sector is completely eliminated -  
``screened'' \cite{scre}.

Let us introduce one heavy neutral  state $S$ for each generation and
consider mass matrix in the basis $(\nu, N^c,  S)$ of the following form 
\be
m = 
\left(\begin{array}{ccc}
0 & m_D & 0\\
m_D^T & 0 & M_D^T\\
0 & M_D & M_S
\end{array}
\right). 
\label{dss}
\ee
Here $M_S$ is the Majorana mass matrix of new fermions.
For $m_D \ll M_D \ll M_S$ it leads to the double (cascade)
seesaw mechanism~\cite{dss}:
\be
m_{\nu} = m_D^T M_D^{-1 T} M_S M_D^{-1 } m_D, 
\label{doubless}
\ee
and $M_R = - M_D M_S^{-1} M_D^{T}$. 
If two Dirac mass matrices are proportional each other,  
\be
M_D = A^{-1} m_D, ~~~~ A \equiv  v_{EW}/V_{GU}, 
\label{propo}
\ee
they cancel in (\ref{dss}) and we obtain
\be
m_{\nu} = A^2 M_S.
\ee
That is, the structure of light neutrino mass matrix is determined by
$M_S$ immediately and does not depend on the Dirac mass matrix.
In this case the seesaw mechanism provides  scale of
neutrino masses but not
the flavor structure of the mass matrix.
It can be shown that at least in SUSY version the  radiative corrections do 
not destroy screening \cite{scre}. The relation (\ref{propo}) can be a consequence 
of Grand Unification with extended gauge group or/and certain 
flavor symmetry~\cite{scre}.   
 
Structure of the light neutrino mass matrix depends now on $M_S$ which can be 
related to some physics at the  Planck scale, and consequently lead to usual 
neutrino properties. In particular,
(i) $M_S$ can be the origin of ``neutrino'' symmetry;
(ii) the matrix  $M_S \propto I$  leads to the quasi-degenerate 
spectrum;
(iii) $M_S$ can be the origin of bi-maximal or maximal mixing
thus leading to the QLC relation 
if the charged lepton mass matrix generates the CKM rotation.


\section{Conclusions}

It may happen that neutrinos prepare new surprise for us on the top of 
existence of large mixings.

It is  difficult to construct complete theory of
quark and lepton  masses. Still we can try 
to answer some generic questions: 

-  Is ``neutrino symmetry'' accidental or not?

-  What are relations between  quarks and leptons 
(universality, symmetry, complementarity)? 

-  Do new neutrino states and their mixing with active neutrinos exist?

In a sense,  we are on the cross-roads and our further advance
may depend on how we will answer these questions.
The way to answer  is precision measurements of neutrino 
parameters
(some benchmarks are identified), study of test equalities, searches
for new (sterile) neutrinos.
It may happen,  that something important (in principles or context)
is still missed.


\end{document}